**A 4-fold-symmetry hexagonal ruthenium for magnetic heterostructures exhibiting enhanced perpendicular magnetic anisotropy and tunnel magnetoresistance**

*Zhenchao Wen\*, Hiroaki Sukegawa, Takao Furubayashi, Jungwoo Koo, Koichiro Inomata, Seiji Mitani, Jason Paul Hadorn, Tadakatsu Ohkubo, and Kazuhiro Hono*

Dr. Z. C. Wen, Dr. H. Sukegawa, Dr. T. Furubayashi, Prof. K. Inomata, Prof. S. Mitani, Dr. J. Hadorn, Prof. T. Ohkubo, Prof. K. Hono

National Institute for Material Science (NIMS)

Tsukuba, 305-0047, Japan

E-mail: Wen.Zhenchao@nims.go.jp

Mr. J. W. Koo, Prof. S. Mitani, Prof. K. Hono

Graduate School of Pure and Applied Sciences, University of Tsukuba,

Tsukuba 305-8577, Japan

Well-designed magnetic heterostructures are indispensable for high-performance spintronic devices, as well as semiconductor heterostructures for a variety of significant information processing devices.[1-3] Engineering a particular crystallographic orientation in multilayer stacks for spintronic devices is a crucial way of achieving ideal magnetic and spin-dependent properties, such as the successfully established fcc(111)-oriented giant magnetoresistance spin valves and (001)-oriented CoFeB/MgO magnetic tunnel junctions (MTJs), which have been utilized for hard disk drive read heads and magnetoresistive random access memories (MRAMs).[4-8] Such a structural control for heterostructures is also of quite importance for the present perpendicular magnetic recording media of hard disk drives (HDDs), which dramatically enhance the areal density of data storage.[9] Moreover, with the trend of scaling down in electronic devices to a few tens of nanometers, the PMA in such a heterostructure is likely to provide a unique way of attaining sufficiently high thermal stability of magnetization for spintronic devices, where conventional materials will be expected to face serious scaling issues.[10-16]

The PMA arising mainly at the interface between ferromagnet (FM) and oxide, due to spin-orbit coupling, is now actively being investigated. It is of particular importance not only for spin transfer torque (STT) MRAM applications using perpendicularly magnetized MTJs (p-MTJs) but also for prospective spintronic devices with electric field or voltage-controlled magnetization switching.[17-21] The perpendicular magnetization of ultrathin films of Fe, fcc-Co, and CoFeB has been realized when combined with oxides such as amorphous $AlO_x$ and crystalline MgO layers, which are attributed to the hybridization of orbitals between ferromagnetic metal and oxygen atoms and the spin-orbit coupling in the ferromagnetic metal atoms.[22-24] In particular, p-MTJs with structures of CoFeB/MgO/CoFeB on a Ta/Ru/Ta composite buffer were demonstrated with an out-of-plane TMR ratio of 124% and a PMA energy density ($K_u$) of $2.1 \times 10^6$ erg cm$^{-3}$ at room temperature (RT) when annealed at 300 °C.[13] However, high magnetic damping of an ultrathin CoFeB layer was observed in the p-MTJs and was reported to be one of the obstacles for further reduction in the current density of STT switching. More recently, the perpendicular magnetization of thin films of $B2$-type Co$_2$FeAl (CFA) alloy was achieved, combining a MgO barrier[18, 25, 26] and a p-MTJ with the structure of CFA/MgO/CoFeB, on a Cr(001) buffer in which an out-of-plane TMR ratio of 91% was obtained.[21] Co$_2$$YZ$ alloys (where $Y$ is a transition metal and $Z$ is a main group element) with $L2_1$ structure, so-called full-Heusler alloys, are attracting increasing interest owing to the high Curie temperature of around 1000 K,[27] high spin polarization,[28-32] and low magnetic damping.[33, 34] For example, CFA is known to possess a very low intrinsic damping constant as low as around 0.001,[33] which is significantly lower than the 0.01 of CoFeB in thick films.[35] In particular, the giant in-plane TMR ratio of 785% at 10 K (360 % at RT) and 1995% at 4.2 K (354% at RT) were achieved in MTJs with CFA and Co$_2$MnSi full-Heusler alloy electrodes, respectively.[29, 32] It was also reported that an optimized buffer layer plays an important role for achieving a higher TMR ratio owing to the improvement in the ordering of full-Heusler alloys and the minimization of lattice misfit or inter-diffusions.[32] Because of the special half-metallic band structures, tunable Fermi level, and low damping constant of Co-based full-Heusler alloys, the well-designed heterostructures with a much higher out-of-plane TMR ratio and enhanced PMA are particularly desirable.

In this study, we report an unusual crystallographic orientation of a Ru layer epitaxially grown on a single crystalline MgO(001) substrate, which is out of common knowledge of epitaxial growth, and furthermore, dramatically enhances the PMA of CFA/MgO interfaces and TMR ratios of full MTJ heterostructures. Very thin monocrystalline CFA(1 nm)/MgO(2 nm) bilayers are grown on a Ru-buffered MgO(001) substrate, which exhibit a large PMA of around four times greater than grown on a conventional Cr(001) buffer. Structural analyses reveal the surprising fact that the hexagonal-close-packed (hcp) Ru layer has a 4-fold symmetry with an approximately Ru(02$\bar{2}$3) crystallographic orientation, which attributes to the lattice matching among MgO, Ru, and CFA. An estimated interfacial PMA of 2.16 erg



cm$^{-2}$, which is greater than 1.3 erg cm$^{-2}$ for CoFeB/MgO[13], is obtained for the Ru/CFA/MgO heterostructure, and an enhanced effective PMA of around $3 \times 10^6$ erg cm$^{-3}$ can be maintained even after annealing at 400 °C. The robustness of PMA at a high annealing temperature up to 400 °C is a key factor for fabricating STT-MRAM, ensuring applicability to the process employed by the semiconductor industry. An out-of-plane TMR ratio of 132% at RT was achieved in a Ru/CFA/MgO/CoFeB MTJ, which is significantly higher than the 91% of CFA-based p-MTJs with a Cr(001) buffer[21].

**Figure 1**a illustrates the unusual crystallographic orientation of a Ru layer epitaxially grown on a MgO(001) substrate, where the Ru($02\bar{2}3$) lattice plane is parallel to MgO(001) surface. The structural properties of Ru(40 nm), and subsequently deposited CFA(20 nm) films, were investigated by X-ray diffraction (XRD). In Figure 1b, in addition to the out-of-plane XRD diffraction peaks from the MgO substrate, the CFA layer shows diffraction peaks of (002) at 31.1° and (004) at 64.9° with the absence of an $L2_1$(111) peak, indicating that the CFA film on the Ru-buffered MgO(001) substrate has a (001) orientation with a $B$2-ordered structure, in which swapping occurs between Fe and Al atoms while Co atoms occupy the regular sites of $L2_1$ structure. Moreover, the diffraction peak at 116.2° indexed as ($02\bar{2}3$) of hcp Ru is shown, yet no other peaks from Ru are detected in the $2\theta$-$\omega$ scan. This suggests the growth of Ru($02\bar{2}3$) orientation on the MgO(001) substrate, as illustrated in Figure 1a. In order to confirm this further, a $\omega$-scan for the Ru plane was carried out, as shown in the inset of Figure 1b, revealing two peaks at ~58°±2.4°. Thus, the Ru($02\bar{2}3$) plane is approximately parallel to MgO(001), whereas the orientation is distributed with the center position tilted from the sample surface by 2.4°.

To investigate the epitaxial relationship between MgO(001) and Ru($02\bar{2}3$), the in-plane XRD diffraction was measured for the 40-nm-Ru layer. In the measurement of the diffraction vector along MgO[100], the peak of the Ru($1\bar{1}01$) plane was clearly observed, whereas a very



weak Ru($\bar{2}$110) peak was also detected, as shown in Figure 1c. For the diffraction vector along MgO[110], the Ru($\bar{2}$110) and Ru(0$\bar{1}$12) planes were also clearly identified, as shown in Figure 1d. The two Ru planes observed in the MgO[100] and MgO[110] directions indicate that the Ru film has structural domains. The positions of the Ru(1$\bar{1}$01), Ru($\bar{2}$110), and Ru(0$\bar{1}$12) planes in an hcp-Ru structure cell are illustrated in Figure S1 (Supporting Information). The insets of Figures 1c, d show $\phi$-scan measurements of each reflection. Clear 4-fold peaks are observed for all the planes, indicating that the Ru film is epitaxial with a 4-fold symmetrized structure on the MgO(001) substrate. Note that a two-peak structure with a split angle of ~9° was observed for the Ru(1$\bar{1}$01) reflections. For the CFA films grown on the Ru surface, precise measurements of the 2$\theta$-$\omega$ scan for the CFA(002) and (004) peaks are shown in Figure 1e (down). The ratio of the integrated intensity of the (002) to the (004) peak for the CFA film was calculated to be ~0.2, which is comparable to that reported for CFA films with a low magnetic damping constant of 0.001[33]. The corresponding ordering parameter $S$ is estimated to be ~0.98, indicating an almost full degree of $B$2 order for the CFA film (Supporting Information). The $\omega$-scan rocking curves are shown in Figure 1e (up). The lines with a large full-width-at-half-maximum (FWHM) of 8.14° for CFA(002) and 5.44° for (004) peaks indicate the curvature of the CFA lattice plane and an imperfect alignment on the Ru buffer. Figure 1f (down) shows the 2$\theta$-$\omega$ scan for the CFA(202) and CFA(404) planes, measured by tilting the sample to an angle of $\chi$ = 45°. The $\phi$-scan for the (202) reflection, as shown in Figure 1f (up), clearly indicates peaks with 4-fold symmetry. The results reveal an epitaxial growth of Ru/CFA bilayers on the MgO(001) substrate.

On the basis of the XRD analysis, a schematic illustration of the lattice relationship amongst MgO, Ru, and CFA is shown in **Figure 2**, where the Ru(02$\bar{2}$3) plane is the surface. Figure 2a indicates the plan view of Ru(02$\bar{2}$3), MgO(001), and CFA(001) planes, with an epitaxial relationship. There are three adjacent Ru(02$\bar{2}$3) planes illustrated, with the atom



arrays of 'Ru1 and Ru2', 'Ru3 and Ru4', and 'Ru5 and Ru6', respectively labeled in the separated lattice planes. Intersections of Ru($1\bar{1}01$), Ru($10\bar{1}\bar{1}$), Ru($0\bar{1}12$), and Ru($\bar{2}110$) planes are illustrated. The in-plane XRD results indicate the relationship MgO[110]//Ru[$\bar{2}110$]. The 4-fold symmetry of the $\phi$-scan is understood by considering that the Ru layer is a mixture of four variants, with the direction of Ru[$\bar{2}110$] along MgO<110>. The Ru($1\bar{1}01$) plane has a dihedral angle of 40.5 °relative to the Ru($\bar{2}110$) plane lying parallel to MgO(110). The peak positions in the $\phi$-scan of Ru($1\bar{1}01$), shown in the inset of Figure 1c, are consistently explained by considering the variants. The atoms in each Ru($02\bar{2}3$) plane distribute as a squared structure, with a lattice constant of 0.270 nm and 0.265 nm between the nearest neighbor atoms. Based on this spacing, the lattice mismatches in the epitaxial structure of MgO(001)/Ru($02\bar{2}3$)/CFA(001) are estimated to be ~−11% (MgO/Ru) and ~8% (Ru/CFA). These mismatches are relatively large; however, because the distance between the adjacent planes of Ru($02\bar{2}3$), 0.0905 nm, is much shorter than that of the lattice planes of MgO(001) or CFA(001), a zigzag flexural surface of a reconstructed Ru($02\bar{2}3$) plane could appear with higher stability and may decrease the lattice mismatches. The side view of the epitaxial relationship of MgO(001)/Ru($02\bar{2}3$)/CFA(001) along the MgO[110] direction is illustrated in Figure 2b. The Ru($03\bar{3}4$) and Ru($0\bar{1}12$) planes are intersected 3.3° and 93.0° from the Ru($02\bar{2}3$) plane, which could induce surface reconstructions to produce a stable Ru surface.

Furthermore, cross-sectional high-resolution transmission electron microscopy (HRTEM) observation was carried out for the heterostructure with the ultrathin CFA film on the Ru layer in order to inspect the quantified crystalline properties. **Figure 3**a shows an HRTEM image of the heterostructure of Ru(40)/CFA(1.2)/MgO(1.8)/Fe(0.1)/CoFeB(1.3)/Ta(5) (unit: nm) annealed at 325 °C, viewed along the [110] direction of the MgO substrate. The vertical and horizontal directions correspond to the [001] and [$\bar{1}10$] axes of the MgO substrate, respectively. The CoFeB refers to $Co_{20}Fe_{60}B_{20}$. The image shows that all of the interfaces are



locally smooth, and that the Ru layer shows epitaxial growth. The inset of Figure 3a shows a nano-beam electron diffraction (NBD) pattern of the Ru buffer layer, which reveals that the ($02\bar{2}3$) plane is approximately parallel to the film stack. From the HRTEM image, dihedral angle between the Ru(0001) and Ru($02\bar{2}3$) planes range from 49 to 51 °, and the average in-plane atom spacing of Ru is evaluated to be around 0.27 nm. These results confirm that the obtained Ru layer is consistent with the model illustrated in Figure 2b. The CFA layer and MgO barrier show (001)-orientated growth in the vertical direction. The average in-plane lattice spacing of the ultrathin CFA layer, i.e., $(100)_{CFA}$, is estimated to be 0.285 nm at a detectable level. For the MgO barrier, the average in-plane lattice spacing is 0.149 nm along the [110] axis, which is almost the same as the bulk MgO(220) lattice spacing. In order to characterize the actual surface of the Ru film, reflection high energy electron diffraction (RHEED) patterns were taken for a sample with a structure of MgO(001)-substrate/Ru(40 nm) along the MgO[100] and MgO[110] azimuths, as shown in Figures 3b, c. A typical domain surface with a distinct 4-fold symmetry and without sign of facet planes is observed. Regarding the Ru thickness dependence of the crystallographic orientation, RHEED and XRD patterns were also taken for 5- and 10-nm-thick Ru films deposited on MgO(001) substrates, as shown in Figure S3 (Supporting Information). The results indicate that the unusual crystallographic orientation has no significant dependence on the Ru thickness. Additional effects, such as annealing temperature effect, might cause a slight modification of the crystallographic structure.

The perpendicular magnetization of CFA full-Heusler alloy thin films on a 4-fold symmetrized Ru layer was achieved when facing an MgO layer. **Figure 4**a shows the out-of-plane magnetization hysteresis (*M-H*) loops for the stacks of MgO(001)-substrate/Ru(40)/CFA($t_{CFA}$)/MgO(2)/Ta(5) (unit: nm), with a nominal thickness of $t_{CFA}$ = 0.8–1.5 nm, annealed at 300 °C. The easy axis deviates clearly from out-of-plane to in-plane with an increase in thickness. The threshold thickness is 1.3 nm, which is greater than 1.1 nm



of CFA thin films on a Cr(001) buffer[21]. The PMA of CFA films observed at the thin regime indicates that the interface perpendicular anisotropy plays a significant role in the effective magnetic anisotropy of the films. The PMA at the interface between the FM and oxide can be explained by the hybridization between Co- or Fe-3$d$ and O-2$p$ electron orbitals at the interface[22-24]. The annealing temperature ($T_{ex}$) dependence of PMA energy density ($K_u$) for 1-nm-thick CFA films on Ru and Cr buffers is shown in Figure 4b. $K_u$ was determined by calculating the difference in the areas between the out-of-plane and in-plane magnetic hysteresis loops.[36] The enhancement of $K_u$ is clearly observed in the sample with the Ru buffer at the annealing temperature range of 250–400 °C. The $K_u$ value of $3.1 \times 10^6$ erg cm$^{-3}$ for 1-nm-CFA is obtained in the heterostructure of Ru/CFA/MgO, which is around 4 times higher than $8.0 \times 10^5$ erg cm$^{-3}$ of the 1-nm-thick CFA with a Cr(001) buffer annealed at $T_{ex}$ = 350 °C. Note that 1-nm-thick CFA is necessary for achieving a high TMR ratio as compared with a thinner CFA film[21]. The out-of-plane and in-plane $M$-$H$ loops of the 1-nm-thick CFA films, grown on Cr and Ru buffers with MgO capping, are shown in Figures. 4c,d, respectively. The significant enhancement of the anisotropy field ($H_k$) of ~7 kOe along the hard axis is clearly observed in the sample with the Ru buffer, indicating a large PMA. The out-of-plane $M$-$H$ curves of the uniform film show the easy saturation and good squareness, demonstrating the establishing of perpendicular magnetic easy axis in the samples, which is typical behavior of PMA materials.[13] At $T_{ex}$ > 350 °C, the magnetic easy-axis of 1-nm-thick CFA with a Cr(001) buffer converts from out-of-plane to in-plane, which can be ascribed to the diffusion of Cr atoms into the interface between CFA and MgO[23]. However, an enhanced $K_u$ value of around $3 \times 10^6$ erg cm$^{-3}$ for the 1-nm-thick CFA on a Ru buffer was maintained even after annealing at $T_{ex}$ = 400 °C, as shown in Figure 4b, which could be attributed to the suppression of inter-diffusion among Ru, CFA and MgO layers because of higher melting



temperature of Ru compared to that of Cr. The high annealing temperature is beneficial for improving the ordering of CFA, thereby achieving a higher TMR ratio.

To further evaluate the magnetic anisotropy of the CFA thin films on the Ru layers, the thickness and annealing temperature dependence was studied in the range of $t_{CFA}$ = 0.6–2.1 nm and $T_{ex}$ = 200–450 °C, as shown in **Figure 5**. The product of $K_u$ and $t_{CFA}$ is plotted as a function of $t_{CFA}$ in Figure 5a for the as-deposited samples and for those annealed at 300 °C and 450 °C, respectively. Decreases in the absolute value $|K_u t_{CFA}|$ are observed in the very thin CFA films of $t_{CFA}$ < ~0.8 nm, which is mainly attributed to a decrease in saturation magnetization $M_s$. A linear dependence of $K_u t_{CFA}$ with $t_{CFA}$ is shown for thicknesses above ~0.9 nm, which can be expressed by

$$K_u \cdot t_{CFA} = (K_v - 2\pi M_s^2) t_{CFA} + K_s, \qquad (1)$$

where $K_v$ and $K_s$ are the volume and interface contributions to the total magnetic anisotropy of the films, respectively. The data were fitted in the linear range between 0.9–1.7 nm by a linear function for estimating $K_v$ and $K_s$. The positive (negative) values represent the out-of-plane (in-plane) magnetic anisotropy. We note that for the thicker CFA films ($t_{CFA}$ > ~1.8 nm), a decline in the slope $K_b = (K_v - 2\pi M_s^2)$ of $K_u \cdot t_{CFA}$ versus $t_{CFA}$ curves is observed, indicating a decrease in $|K_v|$ that could be attributed to the release of tetragonal distortion of the in-plane lattice constant of CFA thin films with an increase in thickness. The in-plane lattice constant of very thin CFA films is determined by the Ru and MgO layers in the epitaxial architecture, but becomes more dependent on the bulk CFA with an increase in $t_{CFA}$. The $T_{ex}$ dependence of $K_s$, $K_b$, and $K_v$ is shown in Figure 5b. The as-deposited samples show in-plane magnetic easy-axis for all thicknesses. An estimated $K_s$ value of 0.22 erg cm$^{-2}$ is obtained, with $K_v$ nearly negligible for the non-annealed samples, indicating that the in-plane magnetic anisotropy is dominated by the shape anisotropy $2\pi M_s^2$. The PMA of the thin CFA films appears after annealing, and the $K_s$ increases with an increase in $T_{ex}$, both of which are attributable to the improved crystallization of the MgO layer and an optimized oxidation at the CFA/MgO interface. The maximum value of $K_s$ = 2.16 erg cm$^{-2}$ is evaluated at $T_{ex}$ = 350 °C, corresponding to the high PMA in Figure 4b, which is significantly higher than the 1.3 erg cm$^{-2}$ for CoFeB/MgO[13]. For further increasing $T_{ex}$, a decrease in $K_s$ is observed, which could be attributed to disorder at the CFA/MgO interface.[23] The $K_s$ value observed in the Ru buffer samples is around two times greater than 1.04 erg cm$^{-2}$ of Cr(001) buffer samples[18], which results in a larger PMA of CFA thin films on the Ru buffer.

In order to clarify the origin of such large interface anisotropy, the contributions of both Ru/CFA and CFA/MgO interfaces need to be taken into account. Sandwiched stacks with a structure of Cr(40)/CFA($t_{CFA}$)/Cr(7) and Ru(40)/CFA($t_{CFA}$)/Ru(7) (unit: nm) were fabricated on MgO(001) substrates with varied $t_{CFA}$ =1.0, 1.2, 1.4, and 1.6 nm. A $K_s$ value of 0.19 and 0.20 erg/cm$^2$ was obtained with structures of Cr/CFA/Cr and Ru/CFA/Ru, respectively, after annealing at $T_{ex}$ = 350 °C. This indicates that the large PMA of CFA thin films in Ru/CFA/MgO mainly results from the contribution by the interface of CFA/MgO, promoted by the Ru buffer layer. Furthermore, the surface morphology of CFA thin films on Cr and Ru buffers was characterized by atomic force microscopy (AFM), and is shown in Figure S4 (Supporting Information). The extremely flat surface of 1-nm-thick CFA on a Cr(001) buffer was achieved with an average surface roughness ($R_a$) of 0.08 nm and a peak-to-valley (*P-V*) value of 0.8 nm, whereas a relatively high surface roughness is observed for CFA on the Ru buffer with $R_a$ = 0.24 nm and *P-V* = 2.7 nm. However, the $R_a$ is still much smaller than the CFA thickness (1 nm). In addition, the HRTEM image in Figure 3a shows a locally smooth CFA/MgO interface. With regard to the origin of the enhanced PMA, we can infer that the improved CFA/MgO interface may contribute to the larger interfacial PMA. The lattice of CFA ultrathin films may be almost unconstrained from the Ru buffer, which leads to an optimized lattice combination between CFA and MgO. However, the CFA lattice with a structure of Cr/CFA/MgO is dominated by the Cr(001) lattice, owing to the very small lattice mismatch (~0.6%) between Cr and CFA, whereas the lattice mismatch between CFA and MgO of ~3.8% results in a less perfect CFA/MgO interface with a subsequent reduction of interfacial PMA. A similar trend of $K_v$ and $K_b$ was found, as shown in Figure 5b. A decrease in $K_v$ is observed with an increase in $T_{ex}$, and the minimum $K_v$ value is around $-9 \times 10^6$ erg cm$^{-3}$ obtained at $T_{ex}$ = 350 °C. The origin of this large negative $K_v$ could be attributed to the tetragonal distortion of the in-plane lattice constant of CFA thin films. The magnitude is ~2 orders larger than $8 \times 10^4$ erg cm$^{-3}$ of the magnetocrystalline anisotropy of the CFA full-Heusler alloy without a tetragonal distortion[33]. Regarding the tetragonal distortion induced magnetic anisotropy change, CFA can have a positive or negative value for the magnetic anisotropy depending on the variation of in-plane lattice constant. From cross-sectional HRTEM images for the system, we obtained that the average in-plane lattice constant of CFA has been expended about 2.5% compared to the bulk $B$2-CFA, which can yield a negative impact on total magnetic anisotropy of the ultrathin CFA layer. Furthermore, the $K_s/t_{CFA(t_{CFA} = 1 \text{ nm})}$ value of $2.16 \times 10^7$ erg cm$^{-3}$ is much higher than $K_v$; thus, the PMA in this structure results from the interfacial perpendicular anisotropy.

Using perpendicularly magnetized CFA full-Heusler thin films with enhanced $K_u$, p-MTJs with the stack of CFA(1.2)/MgO(1.8)/Fe (0.1)/CoFeB(1.3) (unit: nm) were fabricated on a Ru-buffered MgO(001) substrate, and a high TMR ratio of 132% is achieved under an out-of-plane magnetic field at RT. **Figure 6** shows the tunneling resistance of a patterned CFA/MgO/CoFeB p-MTJ as a function of out-of-plane magnetic field ($R$-$H$ loops) at RT and 10 K. The TMR ratio is defined as $100 \times (R_{AP} - R_P)/R_P$, where $R_{AP}$ ($R_P$) is the tunneling resistance of antiparallel, AP (parallel, P) magnetization state between bottom and top electrodes. A TMR ratio of 132% (237%) at RT (10 K), with sharp switching between the P (low resistance) and AP (high resistance) magnetization states, was achieved in the p-MTJ after annealing at $T_{ex}$ = 325 °C. The shapes of the R-H loops indicate perfectly perpendicular magnetization of both the CFA and CoFeB layers of the p-MTJ. The TMR ratio at RT is much higher than 91% of CFA p-MTJs with a conventional Cr(001) buffer[21], which could be attributed to the absence of Cr inter-diffusion and an improved CFA/MgO interface.

In summary, we have demonstrated a novel epitaxial growth of hcp Ru with a 4-fold symmetry and a high crystal index for magnetic heterostructures, which is particularly useful for the growth of perpendicular anisotropy films of a potentially half-metallic Heusler alloys. The unusual orientation and symmetry of the Ru layer attributes lattice matching among MgO, Ru, and CFA, resulting in epitaxial growth in the spintronic heterostructure of MgO/Ru/CFA/MgO and dramatically enhanced PMA and TMR in a full MTJ stack compared to those using a Cr(001) buffer. **Table 1** shows the summary of crystallography between Ru and Cr buffer layers, and PMA and TMR values in their magnetic stacks. The epitaxial architecture based on 4-fold-symmetrized Ru will open a new avenue in the development of engineered heterostructures combining hcp and cubic or tetragonal materials.

**Experimental Section**

All multilayer stacks were deposited by an ultrahigh vacuum magnetron sputtering system, with a base pressure of around $4 \times 10^{-7}$ Pa. The structure of Ru/CFA films on MgO(001) substrates was first characterized by out-of-plane ($2\theta/\omega$-scan) and in-plane ($2\theta_\chi/\phi$-scan) XRD spectra with Cu K$\alpha$ radiation ($\lambda$ = 0.15418 nm). Then, microstructural characterization was performed by TEM (Titan G2 80-200). Thin foil specimen for the TEM observations was prepared by the lift-out technique using a focused ion beam (FIB), FEI Helios Nanolab 650. Stacks with a structure of MgO(001)substrate/Ru(40)/CFA($t_{CFA}$= 0.6−2.1)/MgO(2)/Ta(5) (unit: nm) were annealed at a temperature range of $T_{ex}$ = 200–450 °C in a vacuum furnace for 1 h, in order to investigate the annealing temperature dependence of PMA for CFA thin films on Ru-buffered MgO(001) substrates. The magnetization hysteresis loops under in-plane and out-of-plane magnetic fields were measured at RT using a vibrating sample magnetometer (VSM). The p-MTJ multilayer films were micro-fabricated into junctions with an active area of 5 × 10 μm$^2$ by conventional UV lithography and lift-off technique. The p-MTJs were annealed at a temperature of $T_{ex}$ = 325 ℃ for 1 h, and the magneto-electrical transport properties were measured at RT and low temperature using a dc four-probe method in a physical property measurement system (PPMS).

**Acknowledgements**

This work was partly supported by the Japan Science and Technology Agency, CREST, and JSPSKAKENHI Grant Number 23246006.

Received: ((will be filled in by the editorial staff))
Revised: ((will be filled in by the editorial staff))
Published online: ((will be filled in by the editorial staff))

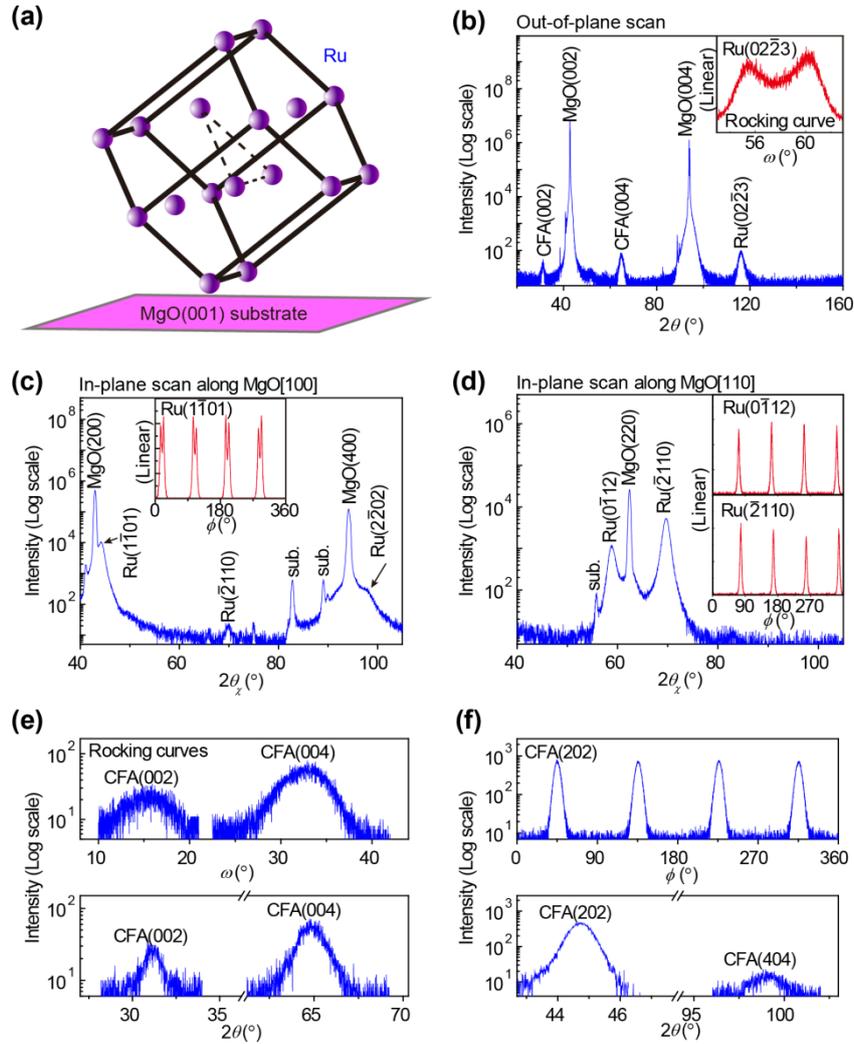

**Figure 1.** (a) Schematic illustration of a (02$\bar{2}$3)-oriented hcp-Ru cell on a MgO(001) substrate. (b) Out-of-plane (2$\theta$-$\omega$ scan) XRD pattern for a specimen with a structure of MgO(001)-substrate/Ru (40 nm)/CFA(20 nm). Inset shows $\omega$-scan for the Ru(02$\bar{2}$3) plane from 53° to 63°. (c, d) In-plane (2$\theta_\chi$/$\phi$-scan) XRD pattern for the structure of MgO(001)-substrate/Ru(40 nm). The scattering vector of the in-plane XRD is parallel to the (**c**) MgO[100] and (d) MgO[110] directions. Insets show the $\phi$-scan XRD pattern for the (**c**) Ru(1$\bar{1}$01) plane and (d) Ru(0$\bar{1}$12) and Ru($\bar{2}$110) planes. (e) Precise 2$\theta$-$\omega$ scan XRD patterns (down) and rocking curves (up) for the peaks of CFA(002) and (004). (f) XRD patterns of 2$\theta$-$\omega$ scan (down) for the CFA film and $\phi$-scan (up) for CFA(202) peak obtained by tilting $\chi$ = 45°.

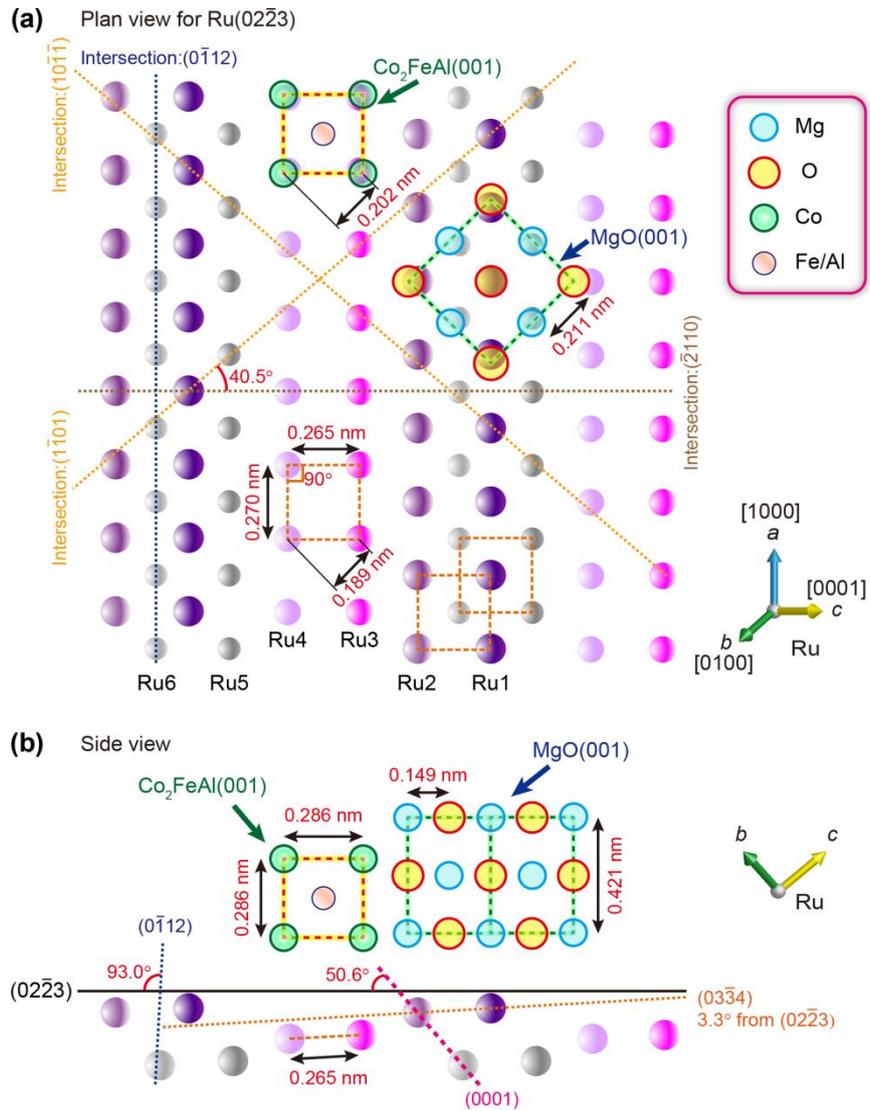

**Figure 2.** (a) Plan view of Ru(02$\bar{2}$3), MgO(001), and CFA(001) planes with an epitaxial relationship. The distribution of Ru atoms on three adjacent Ru(02$\bar{2}$3) planes is illustrated. The squared atom distributions are shown in the planes with lattice constants of 0.270 nm and 0.265 nm, which are matched with MgO and CFA, respectively. (b) Side view of the epitaxial relationship of MgO(001)/Ru/CFA along the MgO[110] direction.

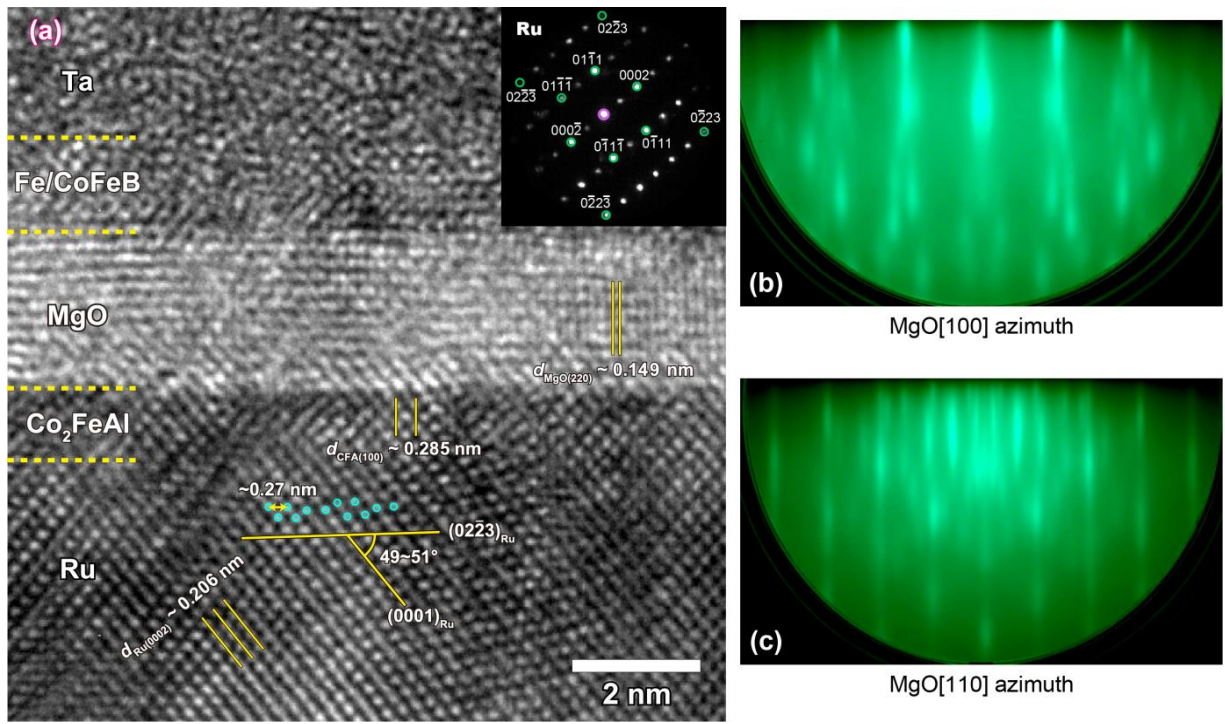

**Figure 3.** (a) HRTEM image for the structure of Ru/CFA/MgO/Fe/CoFeB/Ta on a MgO(001) substrate, viewed along the [110] direction of the substrate. The vertical and horizontal directions correspond to the MgO[001] and [$\bar{1}$10] axis of the substrate, respectively. Note that the last digit in the values of the lattice spacing is not exactly accurate due to the detectable level. Inset shows NBD patterns of the Ru buffer layer. (b, c) RHEED patterns for the surface of 40-nm-thick Ru film on a MgO(001) substrate. The incident electron beam is along the (b) MgO[100] and (c) MgO[110] azimuth.

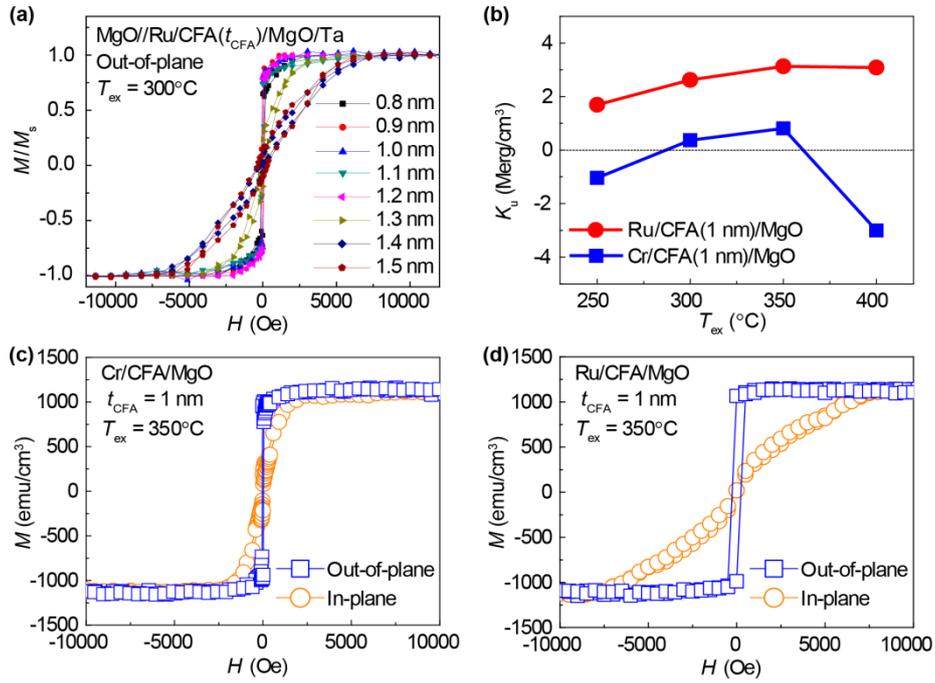

**Figure 4.** (a) Out-of-plane *M-H* loops of CFA ultrathin films with varying thickness on a Ru-buffered MgO(001) substrates. (b) Annealing temperature $T_{ex}$ dependence of PMA energy density $K_u$ for 1-nm-thick CFA with Ru and Cr buffers. (c, d) In-plane and out-of-plane *M-H* loops for 1-nm-thick CFA films, in the structure of Cr/CFA/MgO and Ru/CFA/MgO, respectively.

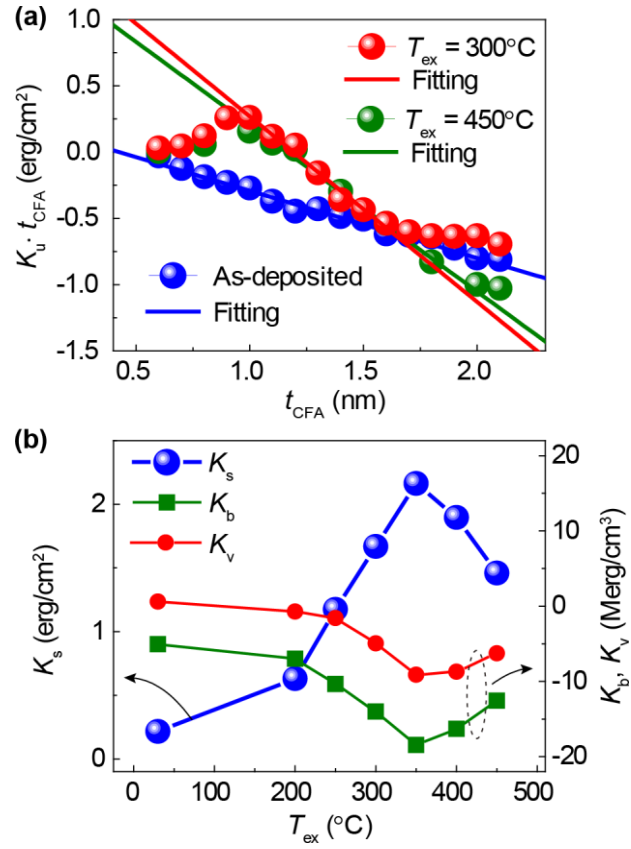

**Figure 5.** (a) The product of $K_u$ and $t_{CFA}$ as a function of CFA thickness for the as-deposited, 300 and 450 °C annealed samples, respectively. The solid lines show linear fitting to the data in a linear data range. $K_s$ and $K_v$ represent interface and volume anisotropy, which can be estimated by the intercept and slope of the fitting curves. (b) Annealing temperature $T_{ex}$ dependence of interface anisotropy $K_s$, slope $K_b$ of $K_u t_{CFA}$ versus $t_{CFA}$, and volume anisotropy $K_v$ in the PMA system.

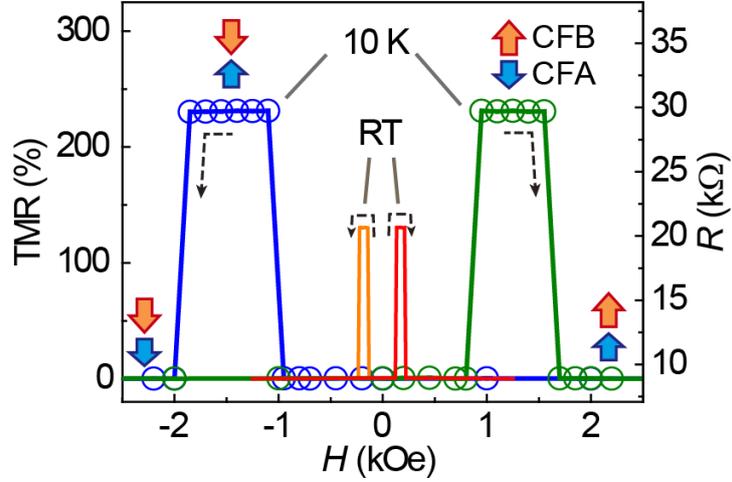

**Figure 6.** Tunneling resistance, $R$, as a function of out-of-plane magnetic field, $H$, measured at RT and 10 K for a patterned p-MTJ with a structure of CFA(1.2)/MgO(1.8)/Fe (0.1)/CoFeB(CFB)(1.3) (unit: nm) on a Ru-buffered MgO(001) substrate. Wide arrows illustrate the magnetization states (P or AP) of bottom and top electrodes. The dashed lines with arrows represent sweeping directions of the magnetic fields with different traces. The directions of the magnetizations of the bottom and top electrodes were determined from $M$-$H$ loops by checking the differences in magnetic moments and switching fields, respectively.

**Table 1.** The comparison of crystallography between Ru and Cr buffer layers, and PMA and TMR values in their magnetic heterostructures.

|  | Ru buffer | Cr buffer |
|---|---|---|
| Crystal Structure | hcp | body-centred cubic (bcc) |
| Orientation | Approximately ($02\bar{2}3$) | (001) |
| PMA value $K_u$ [erg cm$^{-3}$] for 1-nm CFA film | $3.1 \times 10^6$ | $8.0 \times 10^5$ |
| TMR [%] | 132 | 91 |

ToC

An unusual crystallographic orientation of hexagonal Ru with a 4-fold symmetry emerging in epitaxial MgO/Ru/Co$_2$FeAl/MgO heterostructures is reported, in which an approximately Ru(02$\bar{2}$3) growth attributes to the lattice matching among MgO, Ru, and Co$_2$FeAl. Perpendicular magnetic anisotropy of the Co$_2$FeAl/MgO interface is substantially enhanced as compared with those with a Cr(001) layer. The MTJs incorporating this structure gave rise to the largest tunnel magnetoresistance for perpendicular MTJs using low damping Heusler alloys. The 4-fold-symmetry hexagonal Ru arises from an epitaxial growth with an unprecedentedly high crystal index, opening a unique pathway for the development of perpendicular anisotropy films of cubic and tetragonal ferromagnetic alloys.

Keywords: 4-fold-symmetry hexagonal ruthenium; magnetic heterostructures; perpendicular magnetic anisotropy; tunnel magnetoresistance

Zhenchao Wen, Hiroaki Sukegawa, Takao Furubayashi, Jungwoo Koo, Koichiro Inomata, Seiji Mitani, Jason Paul Hadorn, Tadakatsu Ohkubo, and Kazuhiro Hono

A 4-fold-symmetry hexagonal ruthenium for magnetic heterostructures exhibiting enhanced perpendicular magnetic anisotropy and tunnel magnetoresistance

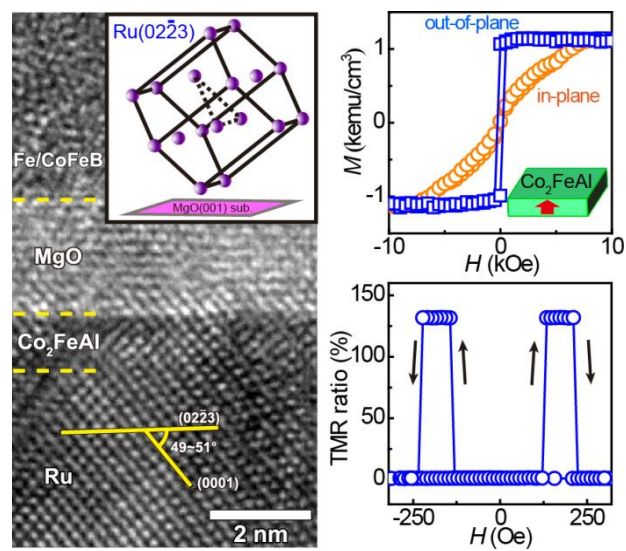

ToC figure